\documentstyle[preprint,eqsecnum,aps]{revtex}
\input epsf       

\newcommand{\MET}{\hbox{$\rlap{\kern0.25em/}E_T$}}
\newcommand{\METcal}{\hbox{$\rlap{\kern0.25em/}E_T^{{\rm cal}}$}}

\begin{document}
\title{Search for $W$ boson pair production in $p\bar{p}$ collisions
at $\sqrt{s}~=~1.8~$~TeV. }
\maketitle
\begin{center}
%
%
%
S.~Abachi,$^{12}$
B.~Abbott,$^{33}$
M.~Abolins,$^{23}$
B.S.~Acharya,$^{40}$
I.~Adam,$^{10}$
D.L.~Adams,$^{34}$
M.~Adams,$^{15}$
S.~Ahn,$^{12}$
H.~Aihara,$^{20}$
J.~Alitti,$^{36}$
% ...
\\
\vskip 0.50cm
\centerline{(D\O\ Collaboration)}
\vskip 0.50cm
\centerline{$^{1}$Universidad de los Andes, Bogota, Colombia}
\centerline{$^{2}$University of Arizona, Tucson, Arizona 85721}
\centerline{$^{3}$Brookhaven National Laboratory, Upton, New York 11973}
\centerline{$^{4}$Brown University, Providence, Rhode Island 02912}
\centerline{$^{5}$University of California, Davis, California 95616}
\centerline{$^{6}$University of California, Irvine, California 92717}
\centerline{$^{7}$University of California, Riverside, California 92521}
\centerline{$^{8}$LAFEX, Centro Brasileiro de Pesquisas F{\'\i}sicas,
                  Rio de Janeiro, Brazil}
\centerline{$^{9}$CINVESTAV, Mexico City, Mexico}
\centerline{$^{10}$Columbia University, New York, New York 10027}
\centerline{$^{11}$Delhi University, Delhi, India 110007}
\centerline{$^{12}$Fermi National Accelerator Laboratory, Batavia,
                   Illinois 60510}
\centerline{$^{13}$Florida State University, Tallahassee, Florida 32306}
\centerline{$^{14}$University of Hawaii, Honolulu, Hawaii 96822}
\centerline{$^{15}$University of Illinois, Chicago, Illinois 60680}
\centerline{$^{16}$Indiana University, Bloomington, Indiana 47405}
\centerline{$^{17}$Iowa State University, Ames, Iowa 50011}
\centerline{$^{18}$Korea University, Seoul, Korea}
\centerline{$^{19}$Kyungsung University, Pusan, Korea}
\centerline{$^{20}$Lawrence Berkeley Laboratory, Berkeley, California 94720}
\centerline{$^{21}$University of Maryland, College Park, Maryland 20742}
\centerline{$^{22}$University of Michigan, Ann Arbor, Michigan 48109}
\centerline{$^{23}$Michigan State University, East Lansing, Michigan 48824}
\centerline{$^{24}$Moscow State University, Moscow, Russia}
\centerline{$^{25}$University of Nebraska, Lincoln, Nebraska 68588}
\centerline{$^{26}$New York University, New York, New York 10003}
\centerline{$^{27}$Northeastern University, Boston, Massachusetts 02115}
\centerline{$^{28}$Northern Illinois University, DeKalb, Illinois 60115}
\centerline{$^{29}$Northwestern University, Evanston, Illinois 60208}
\centerline{$^{30}$University of Notre Dame, Notre Dame, Indiana 46556}
\centerline{$^{31}$University of Panjab, Chandigarh 16-00-14, India}
\centerline{$^{32}$Institute for High Energy Physics, 142-284 Protvino, Russia}
\centerline{$^{33}$Purdue University, West Lafayette, Indiana 47907}
\centerline{$^{34}$Rice University, Houston, Texas 77251}
\centerline{$^{35}$University of Rochester, Rochester, New York 14627}
\centerline{$^{36}$CEA, DAPNIA/Service de Physique des Particules, CE-SACLAY,
                   France}
\centerline{$^{37}$Seoul National University, Seoul, Korea}
\centerline{$^{38}$State University of New York, Stony Brook, New York 11794}
\centerline{$^{39}$SSC Laboratory, Dallas, Texas 75237}
\centerline{$^{40}$Tata Institute of Fundamental Research,
                   Colaba, Bombay 400005, India}
\centerline{$^{41}$University of Texas, Arlington, Texas 76019}
\centerline{$^{42}$Texas A\&M University, College Station, Texas 77843}
%
%
%
%
%
%
%
%
%

\end{center}

\date{\today}

\begin{abstract}
The results of a  search for $W$ boson pair production in $p\bar{p}$
collisions at $\sqrt{s}=1.8$ TeV
with subsequent decay to dilepton ($e\mu, ee $, and $\mu\mu$)  channels
are presented.
One event is observed with an expected background of $0.56\pm0.13$ events
with an integrated luminosity of approximately 14 pb$^{-1}$.
Assuming equal strengths for the $WWZ$ and $WW\gamma$ gauge boson coupling
parameters  $\kappa$  and $\lambda$, limits on the CP-conserving anomalous
coupling constants are $-2.6<\Delta\kappa<2.8$
and $-2.2<\lambda<2.2$ at the 95\% confidence level.
\end{abstract}

\pacs{PACS numbers: ?? 14.80.Er, 13.40.Fn}

\clearpage
The Standard Model (SM) of electroweak interactions makes precise predictions
for the gauge boson self-couplings due to the non-abelian gauge symmetry of
$SU(2)_L\otimes U(1)_Y$.
The $WW\gamma$ coupling has been studied using the cross
section and photon transverse energy spectrum of $W\gamma$ events at UA2
\cite{wgua2},
CDF \cite{wgcdf}, and D\O \cite{wgdzero}.
However, the $WWZ$ trilinear coupling
has not been previously tested. The $W$ boson pair production process
provides a direct test of both the $WW\gamma$ and $WWZ$ couplings \cite{HWZ}.

%

The leading-order SM diagrams for $W$ boson pair production in
$p\bar{p}$ collisions
are $u$- and $t$-channel quark exchange as well as $s$-channel production
with either a photon or a $Z$ boson as the mediating particle. The latter
process
contains the $WW\gamma$ and $WWZ$ trilinear couplings.
The SM predicts that these couplings are
$g_{WW\gamma} = -e$ and $g_{WWZ} = -e\cot{\theta_W}$  and that
unitarity violation due to the $u$- and $t$-channel amplitudes (which depend
on the well-known couplings between the $W$ boson and quarks) is
prevented by cancellations provided by the $s$-channel amplitudes.
Thus, $W$ boson pair production provides a test of the SM gauge structure.


A formalism has been developed to describe the $WW\gamma$ and
$WWZ$ interactions for models beyond the SM \cite{lagrangian}.
The general effective Lorentz invariant Lagrangian
for the electroweak gauge couplings, after imposing C, P, and CP
symmetry, contains six dimensionless coupling parameters:
$g_1^V$, $\kappa_V$, and $\lambda_V$,
where $V =\gamma $ or $ Z$.
$g_1^Z$ is assumed to be equal to $g_1^{\gamma}$,
which is restricted to unity by electromagnetic gauge invariance.
The effective Lagrangian can be reduced to the SM Lagrangian by setting
$\kappa_V=1 \; (\Delta \kappa_V\equiv \kappa_V - 1 =0)$ and
$\lambda_V=0$.
Throughout this letter, it is assumed that
$\kappa_{\gamma} = \kappa_Z$ and
$\lambda_{\gamma} = \lambda_Z$.
The coupling parameters are related to the
magnetic dipole moments ($\mu_W$)
and electric quadrupole moments ($Q_W^e$) of the $W$ boson:
$\mu_W=\frac{e}{2M_W}(1+\kappa+\lambda)$ and
$Q_W^e=-\frac{e}{M_W^2}(\kappa-\lambda)$,
where $e$ and $M_W$ are the charge and the mass of the
$W$ boson \cite{Kim}.

The effective Lagrangian leads to a $W$ boson pair production cross section
which
grows
with $\hat{s}$, the square of the invariant mass of the $WW$ system,
 for non-SM values of the couplings.
In order to avoid unitarity violation, the anomalous couplings are
parameterized as form factors with a scale, $\Lambda$
($e.g.$ $\Delta\kappa/(1+ \hat{s}/\Lambda^2)^2)$.
By requiring that tree-level unitarity is satisfied, a constraint
$\Lambda \leq \left( \frac{6.88}{(\kappa - 1)^2+2\lambda^2 }\right)^{1/4}
\rm{TeV}$ is obtained \cite{HWZ}.
Limits on the coupling parameters $\kappa$ and $\lambda$
are obtained by comparing the measured cross section for
$W$ boson pair production to the predicted non-SM values;
the cross section increases with $\kappa$ and $\lambda$
above the SM prediction of 9.5 pb \cite{ohnemus}.

In this letter the results of a search for
$p\bar{p}(\sqrt{s}=1.8\;{\rm TeV}) \rightarrow WW+X
\rightarrow l\bar{l}'\bar{\nu}\nu '+X$,
where the leptons include muons and electrons, are presented.
The data sample  corresponds to an integrated luminosity of
approximately 14 pb$^{-1}$
collected with the  D\O\ detector
during the 1992-93 Tevatron collider run at Fermilab.

The D{\O} detector\cite{d0nim} consists of three major components:
the calorimeter, tracking, and muon systems.
A hermetic, compensating,
uranium-liquid argon sampling calorimeter with fine transverse and longitudinal
segmentation in projective towers
measures energy out to $ |\eta| \sim 4.0$, where $\eta$ is the
pseudorapidity.
The energy resolution for electrons and photons
is $15\%/\sqrt{E({\rm GeV})}$.
The resolution for the transverse component of missing energy, \METcal , is
$1.1\;{\rm GeV} + 0.02(\sum E_T)$, where $\sum E_T$ is the scalar sum  of
transverse energy, $E_T$,  in GeV, deposited in the calorimeter.
The central and forward drift chambers are used to identify charged tracks
for $|\eta|\leq 3.2.$  There is no central magnetic field.
Muons are identified and their momentum measured with three layers of
proportional drift tubes, one inside and two outside of the magnetized iron
toroids, providing coverage for  $|\eta|\leq 3.3.$
The muon momentum resolution, determined from $J/\psi\rightarrow \mu\mu$ and
$Z\rightarrow \mu\mu $ events,
is $ \sigma (1/p)= 0.18 (p-2)/p^2 \oplus 0.008  $  ($p$ in ${\rm  GeV/c}$).
The $p_T$ of identified muons is used to correct \METcal \ to form the
missing transverse energy, \MET .

Muons are required to be isolated, to have energy deposition
in the calorimeter corresponding to at least that of a minimum ionizing
particle, and to have $|\eta |\leq 1.7$.
For the $\mu \mu$ channel, cosmic rays are rejected by
requiring that the muons have
timing consistent with the beam crossing.
Electrons are identified through the longitudinal and transverse shape of
isolated energy clusters in the calorimeter and by the detection of
a matching track in the drift chambers.  Electrons are required to be
within a fiducial region of $|\eta |\leq 2.5$.
A criterion on ionization $(dE/dx)$, measured in the drift chambers,
is imposed to reduce backgrounds from photon conversions and hadronic
showers with large electromagnetic content.

The event samples come from triggers with dilepton signatures.
The $e\mu$ sample is selected from events passing
the trigger requirement of an electromagnetic
cluster with $E_T\geq 7$ GeV and a
muon with $p_T\geq 5$ GeV/c.  The $ee$ candidates are required to have two
isolated electromagnetic clusters, each with $E_T\geq 10$ GeV.  The $\mu \mu$
candidates are selected from events where at least one muon is identified
with $p_T\geq 5$ GeV/c at the trigger level.

In the offline selection for the $e\mu$ channel,
a muon with $p_T \geq 15$ GeV/c and an electron with $E_T \geq 20$ GeV
are required.
 Both \MET\ and \METcal\ are required to be $\geq 20$  GeV.  In order to
suppress   $Z\rightarrow \tau\bar{\tau}$ and $b\bar{b}$  backgrounds,
it is required that  $20^{\circ} \leq
\Delta\phi(p_T^{\mu},\MET )\leq 160^{\circ}$ if \MET\ $\leq 50$ GeV,
where $\Delta\phi(p_T^{\mu},\MET )$ is the angle
in the transverse plane between the muon and
 \MET. One event survives these selection cuts in a data sample
corresponding to an integrated luminosity of $13.5\pm 1.6$ pb$^{-1}$.

For the $ee$ channel, two electrons are required, each with $E_T \geq 20$ GeV.
The \MET\ is required to be $\geq 20$ GeV.
The $Z$ boson background is reduced by removing events
where the dielectron invariant mass is between $77$ and $105$ GeV/c$^2$.
It is required that $20^{\circ}\leq \Delta\phi(p_T^{e},\MET )\leq 160^{\circ}$
for the
lower energy electron if \MET\ $\leq 50$ GeV.  This selection suppresses
$Z \rightarrow ee$ as well as $\tau\tau$.  The integrated luminosity
in this channel is $13.9\pm 1.7$ pb$^{-1}$.
One event survives these selection requirements.

For the $\mu\mu$ channel, two muons are required, one with $p_T  \geq 20$
GeV/c and another with $p_T  \geq 15$ GeV/c.
In order to remove $Z$ boson events, it is required that the
\MET\ projected on
the dimuon bisector in the transverse plane be greater than $30$ GeV.
This selection requirement is less sensitive to the momentum
resolution of the muons than is a dimuon invariant mass cut. It is
required that
$\Delta \phi (p_T^{\mu},\MET )\leq 170^{\circ}$ for the higher $p_T$ muon.
No events survive these selection requirements in a data sample
corresponding to an integrated luminosity of
$11.8\pm 1.4$ pb$^{-1}$.

Finally, in order to suppress background from $t\bar{t}$ production,
the vector sum of the $E_T$ from hadrons, $\vec{E}_T^{{\rm{had}}}$,
defined as $-(\vec{E}_T^{l1}+\vec{E}_T^{l2}+\vec{\MET })$ is required to
be less than 40 GeV in magnitude for all channels.
Figure 1 shows a Monte Carlo simulation
of $E_T^{{\rm{had}}}$ for $\sim 20$ fb$^{-1}$ of SM $WW$ and $t\bar{t}$ events.
For $WW$ events, non-zero values of $E_T^{{\rm{had}}}$ are due to
gluon radiation and detector resolution.
For $t\bar{t}$ events, the most
significant contribution is the $b$-quark jets from the $t$-quark decays.
This selection reduces the background from $t\bar{t}$ production
by a factor of four for a $t$-quark mass of 160 GeV/c$^2$
and is slightly more effective for a more massive $t$-quark.
The efficiency of this selection criterion for SM $W$ boson pair
production events
is $0.95^{+0.01}_{-0.04}$ and decreases slightly with
increasing $W$ boson pair invariant mass.
The surviving $ee$ candidate passes this selection
requirement but the $e\mu$ candidate \cite{417} is rejected.

The detection efficiency for SM $W$ boson pair production events is
determined using the {\small PYTHIA} \cite{PYTHIA} event generator followed by
a detailed  {\small GEANT}\cite{GEANT} simulation of the D\O \ detector.
Muon trigger and electron identification efficiencies are derived from the
data.
The overall detection efficiency for SM $WW\rightarrow e\mu$ is
$0.092\pm0.010$.  For the $ee$ channel the
efficiency is $0.094\pm0.008$.
For the $\mu\mu$ channel it is $0.033\pm0.003$.
For the three channels combined, the expected number of events for
SM $W$ boson pair production, based on a cross section of
$9.5$ pb \cite{ohnemus}, is $0.46\pm0.08$.
The Monte Carlo program of Ref. \cite{HWZ} followed by a fast detector
simulation \cite{HJthesis} is used to estimate the detection efficiency for
$W$ boson pair production as a
function of the coupling parameters $\lambda$ and $\kappa$.

The backgrounds due to $Z$ boson, Drell-Yan dilepton, $W\gamma$,
 and $t\bar{t}$
events are estimated using the  {\small PYTHIA}
 and {\small ISAJET} \cite{ISAJET} Monte Carlo event
generators followed by the {\small GEANT}
 detector simulation. The backgrounds from
$b\bar{b}$, $c\bar{c}$, multi-jet,
and $W$ + jet events, where a jet is mis-identified as
an electron, are estimated using the data.
The $t\bar{t}$ cross section estimates are from
calculations of Laenen {\it et al.} \cite{topxsec}.
The $t\bar{t}$ background is averaged
for $M_{{\rm top}} = 160$, $170$, and $180$ GeV/c$^2$.
The background estimates are summarized in
Table I.

\begin{table}
\begin{tabular}{lccc}
Background                      & $e\mu$      &  $ee$       &$\mu\mu$ \\ \hline
$Z\rightarrow$ $ee$ or $\mu\mu$ & -----       &$0.02\pm0.01$&$0.066\pm0.026$ \\
$Z\rightarrow \tau \tau$        &$0.11\pm0.05$&$<10^{-3}$   &$<10^{-3}$    \\
Drell-Yan dileptons             & -----       &$<10^{-3}$   &$<10^{-3}$    \\
$W\gamma$                       &$0.04\pm0.03$&$0.02\pm0.01$& -----        \\
QCD                             &$0.07\pm0.07$&$0.15\pm0.08$&$<10^{-3}$    \\
$t\bar{t}$                      &$0.04\pm0.02$&$0.03\pm0.01$&$0.009\pm0.003$ \\
\hline
Total                           &$0.26\pm0.10$&$0.22\pm0.08$&$0.075\pm0.026$ \\
\end{tabular}
\vskip 1.in
\caption{Summary of backgrounds to $WW\rightarrow e e$, $WW\rightarrow
e\mu$ and $WW\rightarrow \mu\mu $. The units are expected number of background
events in the data sample. The uncertainties include both statistical
and systematic contributions.}
\end{table}

The 95\% confidence level upper limit on the $W$ boson pair
production cross section is
estimated  based on one signal event
including a subtraction of the expected background
of $0.56\pm0.13$ events.
The branching ratio $W\rightarrow l\bar{\nu} =  0.108\pm0.004$ \cite{pdgbr}
is assumed.
Poisson-distributed numbers of
events are convoluted with Gaussian uncertainties on the detection
efficiencies, background and luminosity.
For SM $W$ boson pair production, the upper limit for the cross section is
$91$ pb at the 95\% confidence level.
{}From the observed limit, as a function of $\lambda$ and $\kappa$,
 and the theoretical
prediction of the $W$ boson pair production cross section,
the 95\% confidence level limits on
the coupling parameters shown in Fig. 2 (solid line) are obtained.
Also shown in Fig. 2 (dotted line) is the contour of the
unitarity constraint on the  coupling limits for the form factor scale
$\Lambda = 900$ GeV.
This value of $\Lambda$ is chosen so that
the observed coupling limits lie within this ellipse.
The limits on the CP-conserving anomalous coupling parameters are
$-2.6<\Delta\kappa<2.8$ ($\lambda$ = 0) and
$-2.2<\lambda<2.2$ ($\Delta\kappa$ = 0).

The coupling limits are insensitive to the decrease in the expected
$t\bar{t}$ background
which would occur if the top quark is much more massive than
$160-180$ GeV/c$^2$.
If the  top background is negligible, the 95\% confidence level
upper limit
for SM $W$ boson pair production is $93$ pb.

In conclusion, a search for $WW\rightarrow$ dileptons in $p\bar{p}$
collisions at $\sqrt{s}=1.8$ TeV is made.
In approximately 14 pb$^{-1}$
of data, one event is found with an expected
background of
$0.56\pm 0.13$ events.  From the Standard Model, $0.46\pm 0.08$ events
are expected.
For SM $W$ boson pair production, the upper limit for the cross
section is $91$ pb at the 95\% confidence level.
The limits on the CP-conserving anomalous coupling
parameters are $-2.6<\Delta\kappa<2.8$ ($\lambda$ = 0)
and $-2.2<\lambda<2.2$ ($\Delta\kappa$ = 0) at the 95\% confidence level
where $\kappa_{\gamma}$ and $\lambda_{\gamma}$ are assumed to
equal $\kappa_{Z}$ and $\lambda_{Z}$, respectively.
The limits on $\lambda$ and $\Delta \kappa$ exhibit almost no
correlation, in contrast to limits from Refs. [1-3].
The maximum form factor scale accessible for this experiment
is $\Lambda = 900$ GeV.

We thank U.~Baur for providing us with much helpful advice
and D.~Zeppenfeld for the $WW$ Monte Carlo generator and useful instructions.
We thank the Fermilab Accelerator, Computing and Research Divisions,
and the support staffs at the collaborating institutions for their
contributions to the success of this experiment.
We also acknowledge support provided by
the U.S. Department of Energy, the U.S. National Science Foundation,
the Commissariat \`a L'Energie Atomique in France,
the Ministry for Atomic Energy and the Ministry
of Science and Technology Policy in Russia,
CNPq in Brazil,
the Departments of Atomic Energy and Science and Education in India,
Colciencias in Colombia,
CONACyT in Mexico, and
the Ministry of Education, Research Foundation and KOSEF in Korea.


\clearpage
 \vskip 6in
\begin{figure}
 \epsfxsize = 13cm
 \epsffile{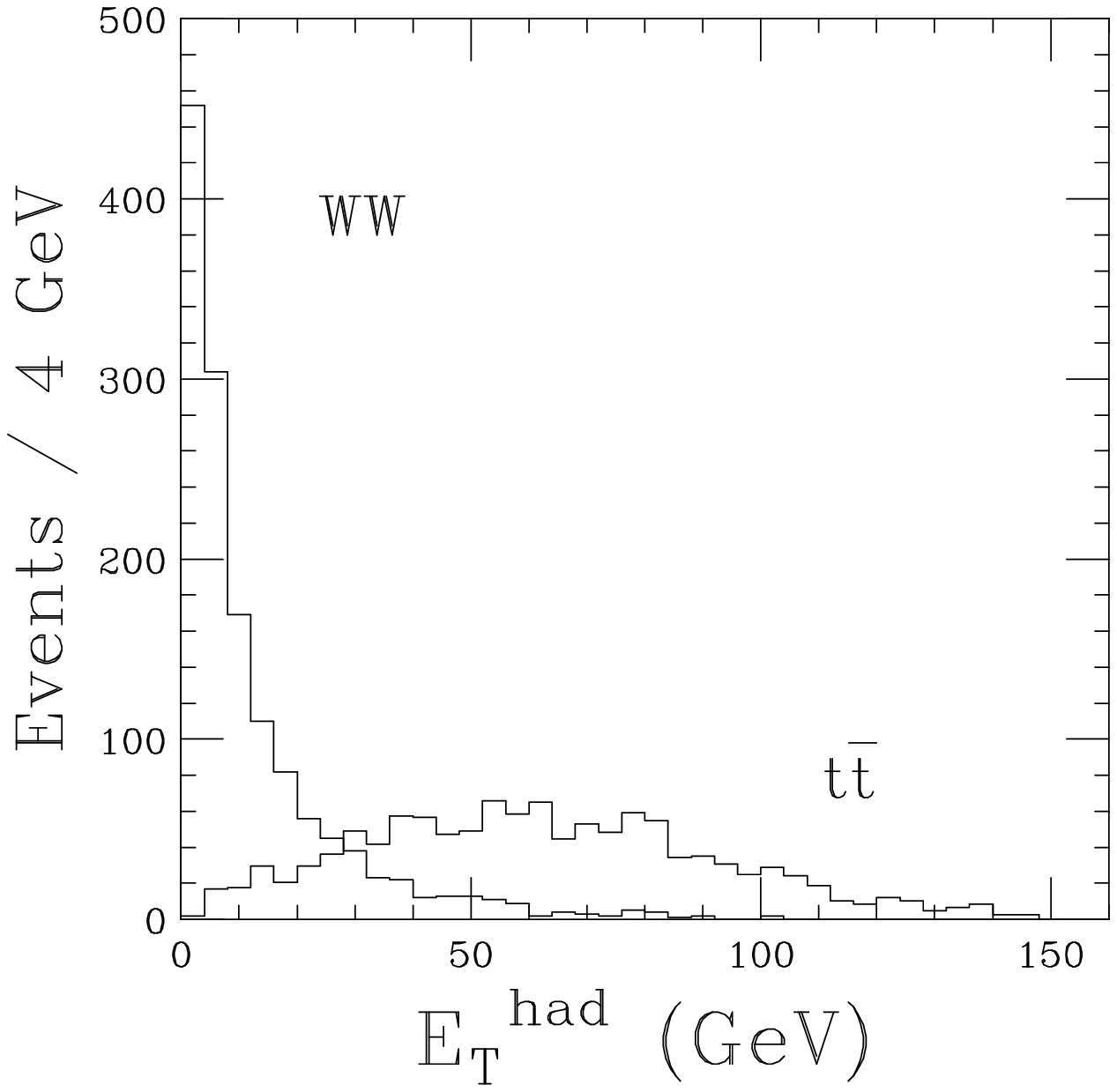}
\vskip 2in
\caption{$E_T^{{\rm{had}}}$ for Monte Carlo
$WW$ and  $t\bar{t}$ events with $M_{{\rm top}} = 160$ GeV/c$^2$
 $(\int Ldt \sim 20$ fb$^{-1})$. Events with $E_T^{{\rm{had}}}
\geq 40 $ GeV were rejected. }
\end{figure}

\clearpage


\vskip 6in
\begin{figure}
\label{contour}
 \epsfxsize = 12.5cm
\centerline{ \epsffile{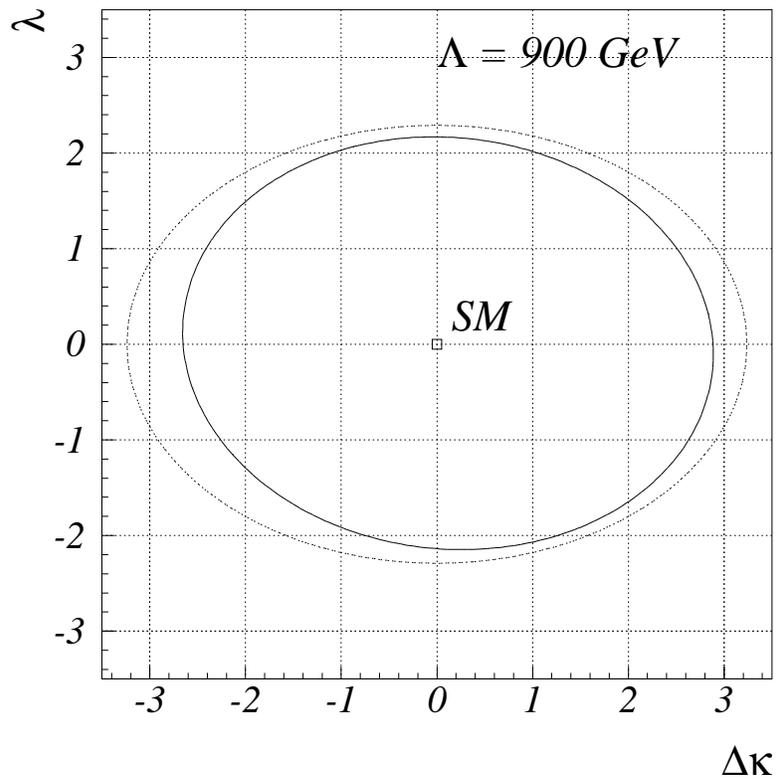} }
\vskip 0.9in
\caption{95\% CL limits on the CP-conserving anomalous couplings $\lambda$ and
$\Delta \kappa$, assuming
that $\lambda_{\gamma} = \lambda_{Z}$ and
$\kappa_{\gamma} = \kappa_Z$.
The dotted contour is the unitarity limit for the form factor scale
$\Lambda = 900$ GeV which was used to set the coupling limits.}
\end{figure}

\end{document}